\documentclass{article}
\usepackage{amsmath}

\input{tcilatex}

\begin{document}

\title{\textbf{An \ Extension of the Van der Waals Equation of State}}
\author{\textbf{Jianxiang Tian}$^{\thanks{%
Email address: lanmanhuayu@yahoo.com.cn}}$ \\
\textit{Department of Physics, Dalian University of Technology }\\
\textit{\ Dalian 116024, P.R.China}\\
\textit{Department of Physics, Qufu }\\
\textit{Normal University, Qufu 273165, P.R.China} \and \textbf{Yuanxing Gui}%
$^{{}}$ \\
\textit{Department of Physics, Dalian University of Technology }\\
\textit{\ Dalian 116024, P.R.China}}
\maketitle

\begin{abstract}
In this paper, we extend the Van der Waals equation of state to a universal
form and find that the Van der Waals equation of state is a special
condition of this form. Then a good form of equation of state for a balanced
liquid-gas coexistence canonical argon-like system is given. The
localization of this universal form is mentioned out in the end.


\end{abstract}

\section{INTRODUCTION}

Much attention has been paid in recent years to the hard core Yukawa (HCY)
potential as a model for the pair interactions of fluids[1]. The liquid
state theories such as the Mean Spherical Approximation (MSA) and the Self
Consistent Ornstein-Zernike Approximation (SCOZA)are proposed. Recent
studies of the HCY fluid can be found in [2,3] and references therein. But
the Van der Waals equation of state%
\begin{equation}
P=\frac{k_{B}T}{v-b}-\frac{a}{v^{2}}
\end{equation}%
was less mentioned recently. Here $a$ and $b$ are parameters, $k_{B}$ is the
Boltzmann constant, and $v=V/N$. In PART 2 of this paper, we extend the VDW
equation of state to a universal form and find that the VDW equation of
state is a special case of this form. In PART 3, we state the localization
of this universal form by applying it to the liquid-gas phase transition to
one order.

\section{Theory\protect\bigskip}

We define particle number density by%
\begin{equation}
n=\frac{N}{V}=\frac{1}{v}.
\end{equation}%
Then Eq.(1) can be expressed as%
\begin{equation}
P=\frac{Nk_{B}T}{V-Nb}-an^{2}.
\end{equation}%
Now we extend the VDW equation of state to such a form as%
\begin{equation}
P=\frac{Nk_{B}T}{V-Nb}-\sigma Bn^{\sigma +1}.
\end{equation}%
$B$ and $\sigma $ are parameters, too. When $\sigma =1$, $B=a$, Eq.(4) reads
the VDW equation of state.\ Now we sign%
\begin{equation}
T^{\ast }=T/T_{c},
\end{equation}%
\begin{equation}
n^{\ast }=n/n_{c},  \label{6}
\end{equation}%
\begin{equation}
P^{\ast }=P/P_{c},  \label{7}
\end{equation}%
with $T_{c}$ the critical temperature , $P_{c}$ the critical pressure , $%
n_{c}$ the critical particle number density.

At critical point, the function $P=P(V,T,N)$ has such qualities as%
\begin{equation}
\frac{\partial P}{\partial V}|_{T_{c}}=0,  \label{8}
\end{equation}%
\begin{equation}
\frac{\partial ^{2}P}{\partial V^{2}}|_{T_{c}}=0.  \label{9}
\end{equation}%
Thus we get the critical data from Eq.(4) by solving Eq.(8) and Eq.(9). They
are%
\begin{equation}
n_{c}=\frac{\sigma }{(\sigma +2)b},  \label{10}
\end{equation}%
\begin{equation}
k_{B}T_{c}=\sigma (\sigma +1)(\frac{2}{\sigma +2})^{2}Bn_{c}^{\sigma },
\label{11}
\end{equation}%
\begin{equation}
P_{c}=\frac{\sigma ^{2}}{\sigma +2}Bn_{c}^{\sigma +1},  \label{12}
\end{equation}%
\begin{equation}
\qquad C=\frac{n_{c}k_{B}T_{c}}{P_{c}}=\frac{4(\sigma +1)}{\sigma (\sigma +2)%
}.
\end{equation}%
Letter $C$ represents the critical coefficient. Substituting Eq.(10-12) to
Eq.(4), we get the reduced equation of state:%
\begin{equation}
P^{\ast }=\frac{4n^{\ast }T^{\ast }\left( \sigma +1\right) }{\left( \left(
\sigma +2\right) -n^{\ast }\sigma \right) \sigma }-\frac{\left( \sigma
+2\right) n^{\ast \left( \sigma +1\right) }}{\sigma }.  \label{14}
\end{equation}%
When $\sigma =1$, Eq.(4) reads%
\begin{equation}
P=\frac{Nk_{B}T}{V-Nb}-Bn^{2},  \label{15}
\end{equation}%
which is just the form of the VDW equation of state. Data in Eq.(10-13) read%
\begin{equation}
n_{c}=\frac{1}{3b},  \label{16}
\end{equation}%
\begin{equation}
k_{B}T_{c}=\frac{8B}{27b},  \label{17}
\end{equation}%
\begin{equation}
P_{c}=\frac{B}{27b^{2}},  \label{18}
\end{equation}%
\begin{equation}
C=\frac{n_{c}k_{B}T_{c}}{P_{c}}=8/3.  \label{19}
\end{equation}%
We are very familiar with these results, which are the data from the VDW
equation of state. Eq.(14) reads%
\begin{equation}
P^{\ast }=\frac{8n^{\ast }T^{\ast }}{\left( 3-n^{\ast }\right) }-3n^{\ast 2},
\label{20}
\end{equation}%
which is the reduced form of the VDW equation of state . From Eq.(3), Eq.(4)
Eq. (14) and Eq.(20), we see that the original form of the VDW equation of
state is%
\begin{equation}
P=\frac{Nk_{B}T}{V-Nb}-1\ast an^{1+1}.  \label{21}
\end{equation}%
When $\sigma =0.7432$, Eq.(4) reads%
\begin{equation}
P=\frac{Nk_{B}T}{V-Nb}-0.7432Bn^{1.7432}.  \label{22}
\end{equation}%
Eq.(10-13) reads%
\begin{equation}
n_{c}=0.2709b^{-1},  \label{23}
\end{equation}%
\begin{equation}
k_{B}T_{c}=0.2609Bb^{-0.7432},  \label{24}
\end{equation}%
\begin{equation}
P_{c}=0.0207Bb^{-1.7432},  \label{25}
\end{equation}%
\begin{equation}
C=\frac{n_{c}k_{B}T_{c}}{P_{c}}=3.4201,  \label{26}
\end{equation}%
Eq.(14) reads%
\begin{equation}
P^{\ast }=\frac{9.3821n^{\ast }T^{\ast }}{\left( 2.7432-0.7432n^{\ast
}\right) }-3.6911n^{\ast 1.7432}.  \label{27}
\end{equation}%
Eq.(26) offers the critical coefficient of argon approximatively.

In 1975, W.G. Hoover, \textit{etc, }generalized the Van der Waals mean-feild
attraction by choosing the attractive potential energy proportional to the $S
$th power of the density[4]. Its per-particle partition function has the
form[4]: \ \ \ \ 
\begin{equation}
z=\frac{V_{e}}{N\lambda ^{3}}\exp [\frac{-\phi _{0}}{Nk_{B}T}-\frac{%
3\varepsilon \phi }{Nk_{B}T}+\frac{B}{k_{B}T}n^{S}].
\end{equation}%
Parametres are interpretated there. $S$ is the exponent of $n.$ We all know
that the equation of state can be obtained by \ \ \ 
\begin{equation}
P=kT\left( {\frac{\partial }{\partial V}\ln z}\right) _{T,N}.
\end{equation}%
Then we can integrate Eq.(4) to get its corresponsive per-particle partition
function. The result is%
\begin{equation}
z_{0}=f(T,N,m)\left( V-Nb\right) ^{N}\exp \left( \frac{B}{k_{B}T}n^{\sigma
}\right) ,
\end{equation}%
with $f$ a function indenpendent with $V.$ Here $m$ is is the mass of one
particle. The equation of state obtained from Eq.(28) is complex. It is
difficult to get its reduced form. Relatively, Eq.(4) is easier to be
operated.

\ \ \ \ If we replace the Van der Waals repulsion term in Eq.(4) with the
Carnahan-Starling hard-sphere repulsion term[5,6], the eqution of state is
written as%
\begin{equation}
P=\frac{Nk_{B}T}{V}\left[ 1+\frac{4y-2y^{2}}{\left( 1-y\right) ^{3}}\right]
-\sigma Bn^{\sigma +1},
\end{equation}%
with $y=g/(4V_{m}).$ Here $g$ is the covolume of Carnahan-Starling
hard-spheres[5], and $V_{m}$ is $mol-$volume. The critical data of Eq.(31)
and the reduced form of Eq.(31) are difficult to be solved. We do not
discuss it in this paper.

\section{\protect\bigskip Discussion}

Here we only study the balanced liquid-gas phase coexistence canonical
system with Eq.(4) and Eq.(14).

In 1945, E.A.Guggenheim collected the data of this system from experiments
and gave out the correlation of $\rho _{1}^{\ast }$, $\rho _{2}^{\ast }$ and 
$T^{\ast }$ by the empirical equations[7] below%
\begin{equation}
\rho _{1}^{\ast }=1+0.75(1-T^{\ast })-1.75(1-T^{\ast })^{1/3},  \label{28}
\end{equation}%
\begin{equation}
\rho _{2}^{\ast }=1+0.75(1-T^{\ast })+1.75(1-T^{\ast })^{1/3}.  \label{29}
\end{equation}%
\bigskip For%
\begin{equation}
\rho _{1}^{\ast }=mn_{1}/mn_{c}=n_{1}/n_{c}=n_{1}^{\ast },
\end{equation}%
\begin{equation}
\rho _{2}^{\ast }=mn_{2}/mn_{c}=n_{2}/n_{c}=n_{2}^{\ast },
\end{equation}%
we have%
\begin{equation}
n_{1}^{\ast }=1+0.75(1-T^{\ast })-1.75(1-T^{\ast })^{1/3},
\end{equation}%
\begin{equation}
n_{2}^{\ast }=1+0.75(1-T^{\ast })+1.75(1-T^{\ast })^{1/3}.
\end{equation}%
In Eq.(32-35), $\rho _{1}^{\ast }$ is the reduced density of the gases and $%
\rho _{2}^{\ast }$ is the reduced density of the liquids. As far as an argon
system is concerned, the inaccuracy of these two equation is generally only
one or two parts per thousand of $\rho _{2}^{\ast }$ or of $\rho _{1}^{\ast
} $ when $T^{\ast }>0.60T_{c}$[7]. So, it is acceptable to consider the data
of $(T^{\ast },n_{1}^{\ast }(T^{\ast }),n_{2}^{\ast }(T^{\ast }))$ from
Eq.(36) and Eq.(37) as the experimental ones in this temperature region when
the balanced liquid-gas coexistence argon-like system is considered. In
Eq.(34-35), $m$ is the mass of one particle. E.A.Guggenheim gave a numerical
analytic result of the relation between the reduced temperature and the
reduced pressure from experiments by equation[8]%
\begin{equation}
P_{e}^{\ast }=\exp (5.29-5.31/T^{\ast }),  \label{34}
\end{equation}%
which best fits the experimental data for argon when $T^{\ast }>0.56T_{c}$
except a tiny region near the critical point[8]. Thus it is acceptable to
consider the data of $(P_{e}^{\ast },T^{\ast })$ from Eq.(38) as the
experimental ones, too. Substituting the data $(T^{\ast },n_{1}^{\ast
}(T^{\ast }))$ from Eq.(36) to Eq.(14), we get the data of the reduced
pressure of the gases when $\sigma $ is fixed. Then we compare these data
with the ones gotten from Eq.(34) to see the accuracy of Eq.(14) when $%
\sigma $ is chosen to be this value. Table.(1) illustrates the data with $%
\sigma =0.7432$, $\sigma =0.8$, $\sigma =1$, and $\sigma =1.5$. $P_{1}^{\ast
}$ is the reduced pressure of the gases gotten by substituting the data $%
(T^{\ast },n_{1}^{\ast }(T^{\ast }))$ in Eq.(36) to Eq.(14). And $%
P_{2}^{\ast }$ is the reduced pressure of the liquids gotten by substituting
the data $(T^{\ast },n_{2}^{\ast }(T^{\ast }))$ in Eq.(37) to Eq.(14). From
Table 1, we can see

\begin{itemize}
\item $|P_{1}^{\ast }-P_{e}^{\ast }|_{\sigma =0.7432}<|P_{1}^{\ast
}-P_{e}^{\ast }|_{\sigma =0.8}<|P_{1}^{\ast }-P_{e}^{\ast }|_{\sigma
=1}<|P_{1}^{\ast }-P_{e}^{\ast }|_{\sigma =1.5}$ when $T^{\ast }>0.8$.

\item $|P_{2}^{\ast }-P_{e}^{\ast }|_{\sigma =0.7432}<|P_{2}^{\ast
}-P_{e}^{\ast }|_{\sigma =0.8}<|P_{2}^{\ast }-P_{e}^{\ast }|_{\sigma
=1}<|P_{2}^{\ast }-P_{e}^{\ast }|_{\sigma =1.5}$ when $T^{\ast }>0.8$.

\item $|P_{1}^{\ast }-P_{e}^{\ast }|_{\sigma T^{\ast }}<|P_{2}^{\ast
}-P_{e}^{\ast }|_{\sigma T^{\ast }}$ when $T^{\ast }$ are fixed in the
region $0.7\leq T^{\ast }\leq 0.99$.
\end{itemize}

So we can conclude that $\sigma =0.7432$ is the best one of these four
values in the temperature region $0>T^{\ast }>0.8$ and Eq.(14) fits the gas
phase better than the liquid phase.

Here we will ask whether a proper $\sigma =\sigma _{0},$ which can bring
right academic forecast to the experimental data $P^{\ast }$, $n_{1}^{\ast }$%
, $n_{2}^{\ast }$ and $T^{\ast }$, exists. Now we suppose that such a proper 
$\sigma _{0}$ exists. Then Eq.(14) is fixed to be the form of%
\begin{equation}
P^{\ast }=\frac{4n^{\ast }T^{\ast }\left( \sigma _{0}+1\right) }{\left(
\left( \sigma _{0}+2\right) -n^{\ast }\sigma _{0}\right) \sigma _{0}}-\frac{%
\left( \sigma _{0}+2\right) n^{\ast \left( \sigma _{0}+1\right) }}{\sigma
_{0}}.  \label{35}
\end{equation}%
Thus we have%
\begin{equation}
y=y(T^{\ast },\sigma _{0})=P^{\ast }(n_{1}^{\ast },T^{\ast },\sigma
_{0})-P^{\ast }(n_{2}^{\ast },T^{\ast },\sigma _{0})=0,  \label{36}
\end{equation}%
where $n_{1}^{\ast }$, $n_{2}^{\ast }$ and $T^{\ast }$ are the experimental
ones in Eq.(36) and Eq.(37). According to our supposion, at two different
reduced temperature $T_{1}^{\ast }$ and $T_{2}^{\ast }$, Eq.(40) will be
true. Fig.(1) is the curve of $y=y(T_{1}^{\ast }=0.8$, $\sigma )$. Fig.(2)
is the curve of $y=y(T_{2}^{\ast }=0.85$, $\sigma )$. From the figures, we
know that $y=y(T_{1}^{\ast }=0.8$, $\sigma )=0$ when $\sigma =\sigma
_{0}(T_{1}^{\ast }=0.8)=1.7050$, $y=y(T_{2}^{\ast }=0.85$, $\sigma )=0$ when 
$\sigma =\sigma _{0}(T_{2}^{\ast }=0.85)=1.9190$. For $\sigma
_{0}(T_{1}^{\ast }=0.8)$ $\neq \sigma _{0}(T_{2}^{\ast }=0.85)$, \textit{the
proper }$\sigma =\sigma _{0}$\textit{\ we try to find does not exist.}

In 1990, J.M.Kincaid, \textit{etc}, representated the pure fluid coexistence
curves by series expansion successfully[9]. Detailed description about pure
fluid coexistence curves such as argon's can be seen there. Our result here
is not as good as theirs, but is more simple in calculation.

\section{Acknoledgement}

This project is supported by the National Natural Science Foundation of
China under Grant No. 10275008. We thank Mrs.Cuihua Zhang, who is working in
the office of Civil administration Sishui, Jining, Shandong Province,
P.R.China, for her help with this paper. We thank Prof. Yunjie-Xia for his
help with the English expression of this paper.

\bigskip

\bigskip \FRAME{ftbpFU}{339.125pt}{257.375pt}{0pt}{\Qcb{The curve of $%
y=y(T_{1}^{\ast }=0.8$, $\protect\sigma )$ versus $\protect\sigma $. The
dotted line is beeline $y=0$ which is a referrence here.}}{}{Figure}{\special%
{language "Scientific Word";type "GRAPHIC";maintain-aspect-ratio
TRUE;display "USEDEF";valid_file "T";width 339.125pt;height 257.375pt;depth
0pt;original-width 513.5pt;original-height 389pt;cropleft "0";croptop
"1";cropright "1";cropbottom "0";tempfilename
'HT3VQ300.wmf';tempfile-properties "XPR";}}\FRAME{ftbpFU}{339.125pt}{%
257.375pt}{0pt}{\Qcb{The curve of $y=y(T_{1}^{\ast }=0.85$, $\protect\sigma %
) $ versus $\protect\sigma $. The dotted line is beeline $y=0$ which is a
referrence here.}}{}{Figure}{\special{language "Scientific Word";type
"GRAPHIC";maintain-aspect-ratio TRUE;display "USEDEF";valid_file "T";width
339.125pt;height 257.375pt;depth 0pt;original-width 513.5pt;original-height
389pt;cropleft "0";croptop "1";cropright "1";cropbottom "0";tempfilename
'HT3VQ401.wmf';tempfile-properties "XPR";}}

$%
\begin{array}{ll}
T^{\ast }=0.7 & 
\begin{array}{lllll}
& \sigma =0.7432 & \sigma =0.8 & \sigma =1 & \sigma =1.5 \\ 
P_{1}^{\ast } & 0.1075 & 0.1042 & 0.0931 & 0.0715 \\ 
P_{2}^{\ast } & -0.5781 & 0.2249 & 5.0082 & -138.7650 \\ 
P_{e}^{\ast } & 0.1007 & 0.1007 & 0.1007 & 0.1007 \\ 
|P_{1}^{\ast }-P_{e}^{\ast }| & 0.0069 & 0.0035 & 0.0076 & 0.0292 \\ 
|P_{2}^{\ast }-P_{e}^{\ast }| & 0.6788 & 0.1243 & 4.9075 & 138.8657%
\end{array}
\\ 
T^{\ast }=0.8 & 
\begin{array}{lllll}
& \sigma =0.7432 & \sigma =0.8 & \sigma =1 & \sigma =1.5 \\ 
P_{1}^{\ast } & 0.2581 & 0.2529 & 0.2339 & 0.1907 \\ 
P_{2}^{\ast } & 0.1786 & 0.5896 & 2.6568 & 32.0708 \\ 
P_{e}^{\ast } & 0.2599 & 0.2599 & 0.2599 & 0.2599 \\ 
|P_{1}^{\ast }-P_{e}^{\ast }| & 0.0018 & 0.0069 & 0.0260 & 0.0692 \\ 
|P_{2}^{\ast }-P_{e}^{\ast }| & 0.0813 & 0.3297 & 2.3969 & 31.8109%
\end{array}
\\ 
T^{\ast }=0.85 & 
\begin{array}{lllll}
& \sigma =0.7432 & \sigma =0.8 & \sigma =1 & \sigma =1.5 \\ 
P_{1}^{\ast } & 0.3681 & 0.3625 & 0.3408 & 0.2876 \\ 
P_{2}^{\ast } & 0.4756 & 0.7420 & 1.9883 & 12.6041 \\ 
P_{e}^{\ast } & 0.3840 & 0.3840 & 0.3840 & 0.3840 \\ 
|P_{1}^{\ast }-P_{e}^{\ast }| & 0.0159 & 0.0215 & 0.0432 & 0.0964 \\ 
|P_{2}^{\ast }-P_{e}^{\ast }| & 0.0916 & 0.3580 & 1.6043 & 12.2201%
\end{array}
\\ 
T^{\ast }=0.9 & 
\begin{array}{lllll}
& \sigma =0.7432 & \sigma =0.8 & \sigma =1 & \sigma =1.5 \\ 
P_{1}^{\ast } & 0.5116 & 0.5061 & 0.4840 & 0.4250 \\ 
P_{2}^{\ast } & 0.7191 & 0.8695 & 1.5264 & 5.5068 \\ 
P_{e}^{\ast } & 0.5434 & 0.5434 & 0.5434 & 0.5434 \\ 
|P_{1}^{\ast }-P_{e}^{\ast }| & 0.0318 & 0.0373 & 0.0594 & 0.1184 \\ 
|P_{2}^{\ast }-P_{e}^{\ast }| & 0.1758 & 0.3261 & 0.9831 & 4.9634%
\end{array}
\\ 
T^{\ast }=0.95 & 
\begin{array}{lllll}
& \sigma =0.7432 & \sigma =0.8 & \sigma =1 & \sigma =1.5 \\ 
P_{1}^{\ast } & 0.7043 & 0.7001 & 0.6821 & 0.6290 \\ 
P_{2}^{\ast } & 0.9035 & 0.9645 & 1.2122 & 2.3443 \\ 
P_{e}^{\ast } & 0.7412 & 0.7412 & 0.7412 & 0.7412 \\ 
|P_{1}^{\ast }-P_{e}^{\ast }| & 0.0369 & 0.0411 & 0.0591 & 0.1122 \\ 
|P_{2}^{\ast }-P_{e}^{\ast }| & 0.1623 & 0.2232 & 0.4710 & 1.6031%
\end{array}
\\ 
T^{\ast }=0.97 & 
\begin{array}{lllll}
& \sigma =0.7432 & \sigma =0.8 & \sigma =1 & \sigma =1.5 \\ 
P_{1}^{\ast } & 0.8028 & 0.7997 & 0.7859 & 0.7428 \\ 
P_{2}^{\ast } & 0.9570 & 0.9896 & 1.1178 & 1.6391 \\ 
P_{e}^{\ast } & 0.8317 & 0.8317 & 0.8317 & 0.8317 \\ 
|P_{1}^{\ast }-P_{e}^{\ast }| & 0.0289 & 0.0321 & 0.0458 & 0.0889 \\ 
|P_{2}^{\ast }-P_{e}^{\ast }| & 0.1253 & 0.1578 & 0.2860 & 0.8073%
\end{array}
\\ 
T^{\ast }=0.99 & 
\begin{array}{lllll}
& \sigma =0.7432 & \sigma =0.8 & \sigma =1 & \sigma =1.5 \\ 
P_{1}^{\ast } & 0.9228 & 0.9214 & 0.9148 & 0.8926 \\ 
P_{2}^{\ast } & 0.9936 & 1.0027 & 1.0370 & 1.1576 \\ 
P_{e}^{\ast } & 0.9290 & 0.9290 & 0.9290 & 0.9290 \\ 
|P_{1}^{\ast }-P_{e}^{\ast }| & 0.0062 & 0.0076 & 0.0142 & 0.0364 \\ 
|P_{2}^{\ast }-P_{e}^{\ast }| & 0.0646 & 0.0737 & 0.1080 & 0.2286%
\end{array}%
\end{array}%
$

\begin{equation*}
Table.1:\text{Data}
\end{equation*}


\begin{thebibliography}{9}
\bibitem{1} Y. Rosenfeld,\textit{\ J. Chem. Phys}. 98, 8126 (1993).

\bibitem{2} C. Caccamo, G. Giunta and G. Malescio, \textit{Mol. Phys.} 84,
125 (1995).

\bibitem{3} D. Pini, G. Stell and N. B. Wilding, \textit{Mol. Phys}. 95, 483
(1998).

\bibitem{4} W. G. Hoover, G. Stell, E. Goldmark and G. D. Degani., \textit{%
J. Chem Phys}. 63, 5434 (1975).

\bibitem{5} N. F. Carnahan, K. E. Starling, \textit{J. Chem. Phys.} 51, 635,
(1969).

\bibitem{6} N. F. Carnahan, K. E. Starling, \textit{AIChE Journal}, 18,
1184, (1972).

\bibitem{7} E.A.Guggenheim, \textit{J. Chem. Phys}, Vol.13, Num.7 (1945).

\bibitem{8} E.A.Guggenheim, \textit{Thermodynamics}, North-Holland Physics
Publishing, 138-139 (1967).

\bibitem{9} J. M. Kincaid, B. Tooker and G. Stell, \textit{Fluid Phase
Equilibria}, Vol.60, Issu.3, 239, (1990).
\end{thebibliography}
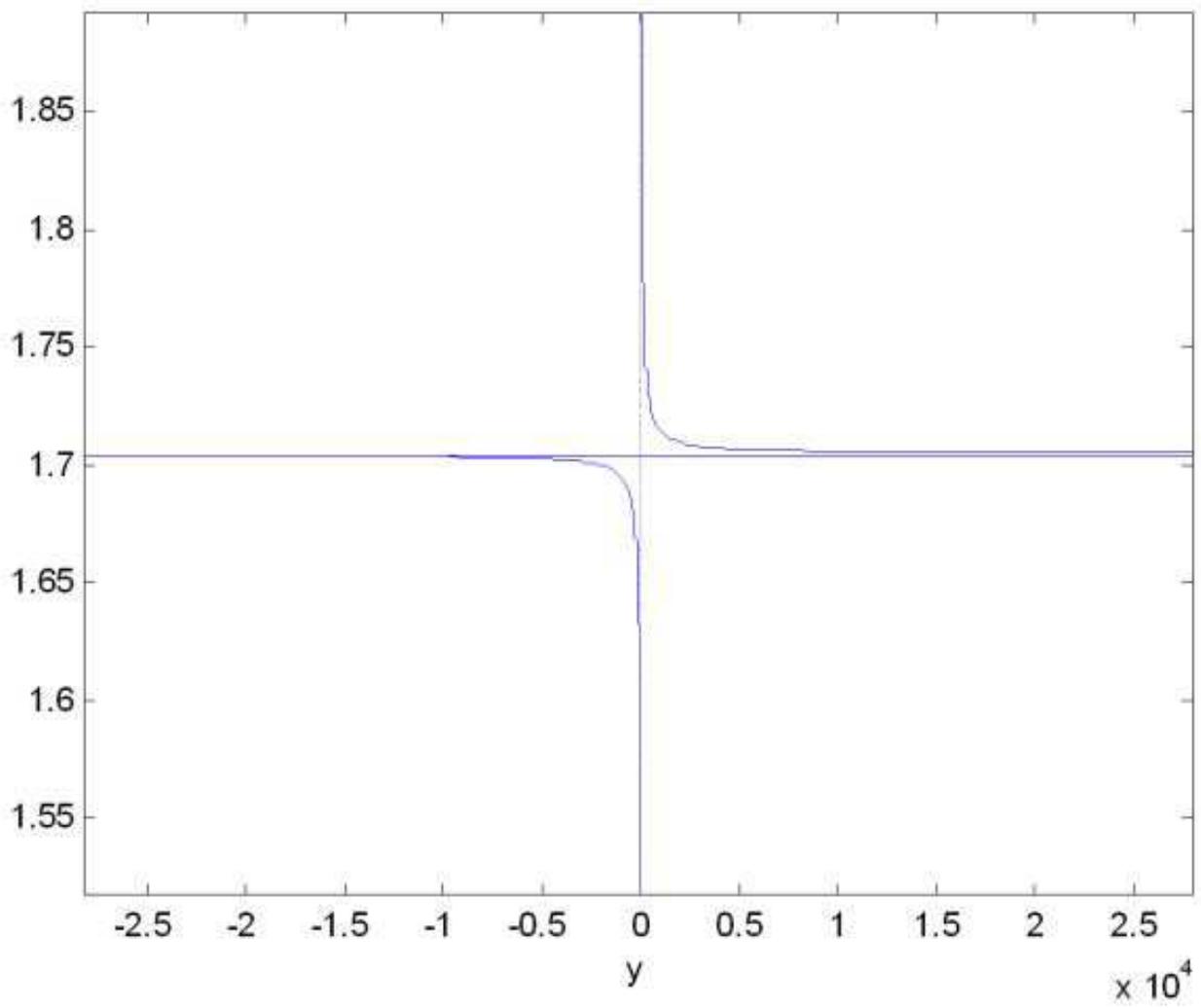
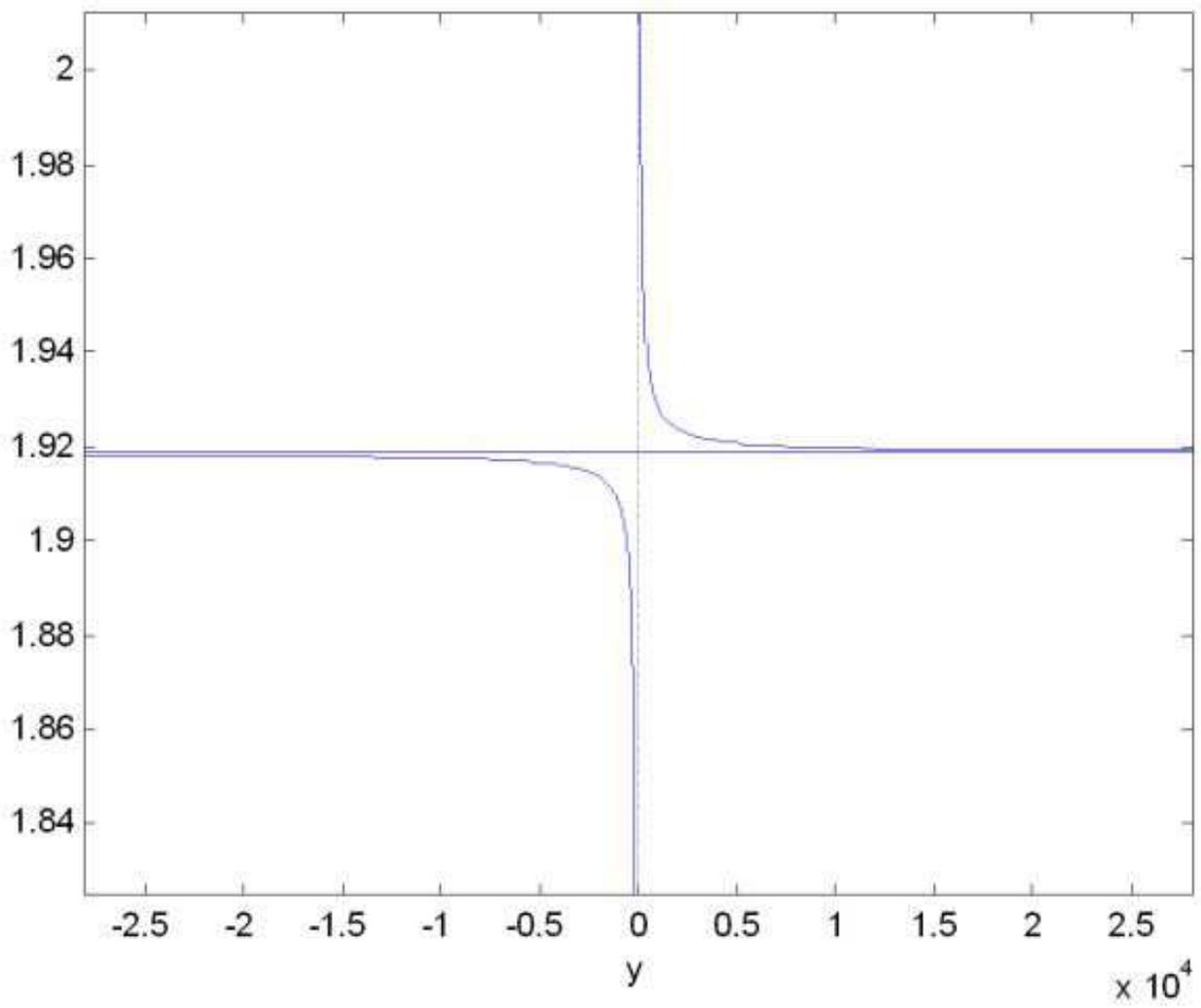
\end{document}